\documentclass[aps,prl,nofootinbib,floatfix,twocolumn]{revtex4}
\usepackage{amssymb,color,graphicx,bm}

\definecolor{red}{rgb}{0.8,0,0}
\definecolor{RED}{rgb}{0.8,0,0}
\definecolor{violet}{rgb}{0.4,0,0.4}
\definecolor{green}{rgb}{0,0.5,0.0}
\definecolor{GREEN}{rgb}{0,0.5,0.0}
\definecolor{navy}{rgb}{0.0,0.0,0.6}
\definecolor{orange}{rgb}{0.8,0.2,0.0}
\definecolor{blue}{rgb}{0.3,0.0,0.8}

\begin{document}

\title{Effect of dynamical noncommutativity on the limiting mass of white dwarfs}

\author{Sayan Kumar Pal, Partha Nandi\\
 S. N. Bose National Centre for Basic Sciences, Salt Lake, Kolkata 700106, India\\ sayankpal@bose.res.in, parthanandi@bose.res.in \\
}

\begin{abstract}
The discovery of the existence of an upper bound on the mass of a white dwarf star is considered as one of the finest of twentieth century astrophysics. On approaching this limiting mass of $1.4M_\odot$, known as the Chandrasekhar mass-limit, 
a white dwarf is believed to spark off with an explosion called type~Ia supernova, which is considered to be a
standard candle. But since the last decade, observations of several 
over-luminous, peculiar type~Ia supernovae indicate that the Chandrasekhar-limit be significantly larger. By postulating noncommutativity between the components of momentum variables, we show that the mass of white dwarfs can be enhanced significantly and 
arrive at a new super-Chandrasekhar mass limit of about $4.68M_\odot$. 
This can provide a plausible explanation for the origin of over-luminous peculiar type~Ia supernovae.
\end{abstract}

\maketitle


\section{Introduction}
White dwarf stars are one of the three end fates of a star. These stars possess a theoretical upper bound on their masses. This fact has been proven repeatedly by numerous observations over the years since its uncovering. The mass-limit of white dwarfs plays an important role
in establishing the type Ia supernovae (SNeIa) to be standard candle, while an SNeIa is expected to be triggered
once a white dwarf approaches its limiting mass, which is $1.4M_\odot$ called the Chandrasekhar 
mass-limit \cite{chandra}. Based on the idea of this standard candle, the accelerated expansion
of universe was established \cite{perlmutter,riess}. However, in last one decade or so,
there are some peculiar, over-luminous SNeIa observed, starting from 2006, whose
high luminosity, given their observed low ejecta velocity, argues their 
progenitor mass to be as high as $2.8M_\odot$, double the Chandrasekhar mass-limit!
Examples of such peculiar over-luminous SNeIa are SN 2003fg, SN 2006gz, SN 200 7if, SN 2009dc 
\cite{howell,scal} and so far at least a dozen of them have been observed. How to explain such a
significant violation of Chandrasekhar-limit? Earlier, proposals were put forth considering enormous efficiency of magnetic fields and it was argued that those highly super-Chandrasekhar
progenitors of SNeIa are plausibly highly magnetized white dwarfs \cite{prl,sub}. 
However, these proposals suffer from the problem of a lack of prior knowledge of field profile within stars and also 
there arises stability questions for such highly 
magnetized white dwarfs\cite{proceed, bera}.

At this circumstance, here we take a stoke of plausible existence of noncommutativity 
of fundamental momentum variables and investigate whether that leads to super-Chandrasekhar
white dwarf with an enhanced limiting mass. The plan is to consider noncommutativity
at the high density regime of non-magnetized white dwarfs. 
The said noncommutativity,
apart from introducing uncertainty in simultaneous measurements of components of momentum variables,
mainly affects the underlying statistical mechanics of electron degenerate matter. 
Note that white dwarf matter is essentially a degenerate electron gas and the pressure 
opposing gravity is mainly the degenerate electron gas pressure throughout. Therefore,
our plan is to explore the effect of momenta noncommutativity in the equation of state (EoS) and 
subsequently the effect of modified EoS to the white dwarf mass.

There are proposals \cite{dop,Born,sny} that at a very high energy regime,
e.g. in early universe, the usual space-space and momentum-momentum
commutativities break down, so that one postulates a NC algebra between the coordinates which are now elevated to the level of operators, 
and similarly among momentum operators.
 Indeed strong plausibility arguments were put forward by Doplicher et. al. \cite{dop} suggesting that a quantum structure of space-time is indispensable at the smallest scales, viz Planck scale. Actually the idea of NC space-times dating back to roughly 1940's \cite{Born, sny} grew as a possible way of removing the infinities which were occurring in quantum electrodynamics at that time. Soon the renormalization program came and the idea of noncommutativity was washed out. However, it is worth mentioning here that almost all candidate theories of quantum gravity currently indicate at this quantum structure of space-time at the fundamental scales \cite{Witten,loop}.
 Actually, the point that all candidate theories of quantum gravity indicating the existence of a quantum structure of space-time is to be now stressed 
in the present work. 
In \cite{Born}, Max Born, in his quest for unifying quantum theory and gravity suggested that both the coordinate space and momentum space should be subjected to similar geometrical laws, as such a Riemannian structure. This is the famous ``principle of reciprocity".
Now, it is a matter of fact that curvature in position space leads to a noncommutativity in the momentum variables. Analogously this implies, curved momentum spaces can lead to noncommutativity among the position variables and all these concepts were beautifully stitched out in a seminal work \cite{Shahn} by S. Majid to understand the nature of quantum gravity using Born's ``principle of reciprocity". It was summed up there, and subsequently more transparently in \cite{Shahn2}, that quantum phase space should contain quantum spacetime and there should be noncommutativity both in the configuration space variables as well as in momentum variables in order to describe quantum gravity. 

White dwarfs have typical densities in the range of $10^6-10^{10} g/cc$ and hence the associated length scales are very far from that of the quantum gravity scale. Therefore one can ignore spatial noncommutativities in the present formulation and we are motivated to figure out the consequences of momentum-momentum noncommutativity arising from curved position space in the 
matter of white dwarf stars, even if not being in the quantum gravity regime. We shall also ignore the inter-particle Coulomb interaction in a white dwarf star following the arguments in \cite{Landau} for simplicity of our present treatment.


\section{Formalism}


We consider relativistic electrons of mass $m_e$ moving in a three-dimensional space where the momentum coordinates $\hat{q}_1$, $\hat{q}_2$ and $\hat{q}_3$ do not commute but the position coordinates $y_i$ commute such that
\begin{eqnarray}
\label{NCHA0}
\left[\hat{q}_{i},\hat{q}_{j}\right]=i\eta_{ij}~,~\left[\hat{y}_{i},\hat{y}_{j}\right]=0~,~\left[\hat{y}_{i},\hat{q}_{j}\right]=i\hbar \delta_{ij}~;~ i,j=1,2,3. \nonumber ~~
\end{eqnarray}
where $\eta_{ij}$ is a $3 \times 3$ anti-symmetric matrix. 
Now we know that any odd antisymmetric matrix can be brought to a block diagonal form wherein  
the bottom right block will be a null matrix. This will render the third momentum coordinate to commute with the other two \cite{spa}. Therefore now we have the NC Heisenberg algebra (NCHA) to be 
\begin{eqnarray}
\label{NCHA}
&&\left[\hat{x}_j,\hat{x}_k\right]=0,~\left[\hat{x}_{j},\hat{p}_{k}\right] = i\hbar\delta_{jk} \nonumber \\ &&
\left[\hat{p}_{a},\hat{p}_{b}\right]=i\eta \epsilon_{ab}, ~~
\left[\hat{p}_a,\hat{p}_z\right]=0,{\rm for}~ a,b=1,2.
\end{eqnarray}
where subscripts $1$ and $2$ respectively imply $x-$ and $y-$components of respective variables. Here $\eta$ is the momentum NC parameter. 
It will turn out to be transparent below that the spectrum will be unaffected 
by such a transformation of the operators, as the spectrum depends on the 
eigenvalues of the $\eta$ matrix and such a kind of rotation does not change 
the eigenvalues of the $\eta$ matrix.

The NC algebra (\ref{NCHA}) we are dealing here is rather based on a physical picture and the origin of noncommutativity in momentum space can be argued heuristically in the following way. Basically, this stems from the
fact that here we are considering the motion of a typical ultra-relativistic electron in the background of the gravitational field produced by the stellar
material, rather than ignoring it, unlike what was done by Chandrasekhar who considered the electrons to be essentially free. But like his analysis, we too are assuming that the potential to be infinitely large outside the star so that the electrons are essentially trapped inside. Let us consider the gravitational interaction on an electron due to the stellar matter which has the effect of creating a curvature in a small region of the ambient space. Now we model a test electron propagating under this gravitational background, i.e. propagating on the associated curved space \cite{Feynman}. 
In our analysis we will be primarily interested in the dynamics of ultra-relativistic electrons, which corresponds to the existence of limiting mass of white dwarfs in the usual case, and for such ultra-relativistic electrons gravity effects are encoded in spatial curvature \cite{Schutz},\cite{dad}. 
Generically speaking, the situation can also be conceived of as the stable circular motion of a test electron under the influence of gravity of the surrounding matter and such a motion can always be mapped to the motion of a free electron on the sphere as geodesic flows on the sphere is equivalent to the Kepler problem \cite{Moser}. This is precisely the case we consider here. Our approach closely tallies with the well-known observations made in \cite{Town} in which the author inferred that there must be strong curvature at short length scales, which, averages to zero at large scales. Bound stable circular orbits for photons under general relativistic gravitational field have been demonstrated in \cite{Dad2}.
All of these facts motivate us to consider effective motion of electrons on the surface of sphere resulting in consideration of noncommutativity of momentum components in the two directions and a commutative momentum for the third direction which is outlined below:

The commutator of the covariant derivatives acting on a fermion (a typical spinor) written in the local orthonormal basis reads as \cite{Car, hehl} -
\begin{equation}\nonumber
\left[\nabla_a, \nabla_b \right]\psi= \frac{1}{4}R_{abcd} \gamma^c \gamma^d ~\psi,
\end{equation}
where $\psi$ is the fermion wave function, $\gamma^{c,d}$ are 
Dirac $\gamma$-matrices and $R_{abcd}$ is the Riemann tensor in the local coordinate.
For a space with constant scalar curvature $R$ ($\frac{2}{r^2}$ for a sphere $S^2$), $R_{abcd}$ is actually a constant and $\gamma$'s are constant matrices. 
If we now define $p_a=-i\hbar\nabla_a$, we obtain
\begin{equation}\label{cur}
\left[p_a, p_b \right]\psi=i\frac{\hbar^2}{4 r^2} S_{ab} ~\psi
\end{equation}
where $S_{ab}=i[\gamma_a, \gamma_b]$.
This is an example of canonical type of commutation relation and this kind of noncommutativity can be said to be spin dependent noncommutativity. Such NC structures have been already encountered in the literature \cite{A. Deri}, \cite{Pop}.
Comparing equations (\ref{NCHA}) and (\ref{cur}), we see that the momentum noncommutativity parameter $\eta$ is mimicking like $\frac{\hbar^2}{4r^2}$.\\

Also recently, similar NC structures among the phase space variables were derived as an effective description from Berry curvature in condensed matter systems like in semiclassical dynamics of Bloch electrons, quantum Hall systems, Fermi liquids, ferromagnetic metals \cite{xiao,hiroaki,son, maslanka}. Further the postulation of noncommutativity in momentum space, which clearly has to stem from the system inhabiting it, has also been addressed along quite similar lines in \cite{thak}. 
Since this is noncommutativity in the momentum space, we have called it as dynamical noncommutativity.
Particularly, the momentum space NC operation generally arises in the presence of a background magnetic field.

\section{Energy spectrum}
Given that we have argued that the momentum NC is
essentially the curvature effect of the background space
along the lines of \cite{Shahn, Shahn2, Town}, we can study the motion of an ultra-
relativistic electron in such a background with only two
components of the momentum being NC in nature.
From here on, we just focus on solving the spectrum of an
ultra-relativistic electron satisfying the algebra (\ref{NCHA}). The standard approach in the literature to deal with such NC quantum mechanical problems is to form an equivalent commutative 
description of the NC theory by employing some non-canonical transformation, the so-called Bopp shift, which relates the NC operators $\hat{x}_{j}$, $\hat{p}_{j}$ following equation (\ref{NCHA}) to ordinary commutative operators $x_{j}$, $p_{j}$, 
satisfying the usual Heisenberg algebra 
\begin{eqnarray}
\left[x_{j}, p_{k}\right]=i\hbar \delta_{jk}~,~~~ \left[p_{1},p_{2}\right]=0.
\label{HA}
\end{eqnarray}
From here onwards, we denote NC operators
 with the hat notation and commutative operators without hat. 
We shall use the following well-known generalised Bopp-shift transformations \cite{Moyal} given as
 
\begin{eqnarray} \nonumber
&&\hat{p}_{j}  =  p_{j}+\frac{\eta}{2\hbar}\epsilon_{jk}x_{k}.\nonumber \\
\end{eqnarray}
If the total Hamiltonian of the system is $\hat{H}$,
the Dirac equation for an electron moving in the NC plane satisfying the NCHA reads
\begin{equation}
\hat{H} \psi =i\hbar\frac{\partial \psi}{\partial t} = E \psi,
\end{equation}
where $\psi$ is a two-component spinor 
of components $\phi$ and $\chi$,
and the Dirac Hamiltonian is given by 
\begin{equation}
\hat{H}=\vec{\alpha}.\vec{p}c+ \beta m_e c^2,
\end{equation}
where $c$ is speed of light.
The above gives a pair of equations
\begin{eqnarray}
(E-m_ec^2)\phi=\vec{\sigma}.\vec{p}c~ \chi \,\,
{\rm and}\,\,
(E+m_ec^2)\chi=\vec{\sigma}.\vec{p}c~ \phi.
\end{eqnarray}
On combining them, we obtain the energy given by
\begin{eqnarray}
\label{energy}
&& (E^2-m_e^2c^4) = (\vec{\sigma}.\vec{p})^2c^2 
= \hat{p}^2c^2 + i \vec{\sigma}.(\hat{p}\times\hat{p})c^2 \nonumber \\
&=& \left[(p_x^2 + p_y^2)+ B(x^2+y^2) +  \frac{\eta}{\hbar}(y p_x -x p_y) +   p_z^2 -\sigma_z \eta\right]c^2 \nonumber \\
\label{eigen}
\end{eqnarray}
where 
\begin{eqnarray} \label{cons}
B=\frac{\eta^2}{4\hbar^2}  \nonumber .
\end{eqnarray}

Therefore, we obtain an equivalent commutative Hamiltonian in terms of the commutative variables 
(quantum mechanical operators) which describes the original system defined over the NC plane.



Now to compute the spectrum of a charged particle in such a NC spacetime, first of all we need 
to construct the ladder operators which will diagonalize the following part 
of right hand side of equation (\ref{eigen}),
given by
\begin{eqnarray}
\hat{H}^{\prime}&=& \left[(p_x^2 + p_y^2)+ B(x^2+y^2)
+ \frac{\eta}{\hbar}(y p_x -x p_y)  \right]c^2.\nonumber \\ 
\label{diag}
\end{eqnarray} 
The ladder operators involving the commutative phase-space variables (operators)
$x, y, p_{x}, p_{y}$, given by 
\begin{eqnarray}
\label{e30a}
a_{j}=\left(\frac{1}{2\hbar\sqrt{B}}\right)^{\frac{1}{2}} \Bigg(p_j-i\sqrt{B}x_j\Bigg), \nonumber \\ a_{j}^{\dagger}=\left(\frac{1}{2\hbar\sqrt{B}}\right)^{\frac{1}{2}} \Bigg(p_j+i\sqrt{B}x_j\Bigg),\nonumber \\
\end{eqnarray}
satisfy the commutation relations 
\begin{eqnarray}
\label{e30}
[a_{x},a_{x}^{\dagger}]=1=[a_{y},a_{y}^{\dagger}].
\end{eqnarray}
Further defining a pair of operators
\begin{eqnarray}
\label{lad}
\hat{a}_{1}=\frac{a_{x}+ia_{y}}{\sqrt{2}},\quad \hat{a}_{2}=\frac{a_{x}-ia_{y}}{\sqrt{2}},
\end{eqnarray}
which satisfy the commutation relations
\begin{eqnarray}
\label{e32a} 
[\hat{a}_{1},\hat{a}_{1}^{\dagger}]=1=[\hat{a}_{2},\hat{a}_{2}^{\dagger}],
\end{eqnarray}
the Hamiltonian given by equation (\ref{diag}) can be recast in the diagonal form as
\begin{eqnarray}
\label{diagfinal}
\hat{H}^{\prime}=\eta (2\hat{a}_{1}^{\dagger} \hat{a}_{1} +1)c^2.
\end{eqnarray}
Therefore, on using equations (\ref{energy}) and (\ref{diagfinal}), the total energy of the system is given by
\begin{equation}
E^2(m)=p_{z}^2(m) c^2 + m_e^2c^4 +2 m \eta c^2,
\label{spectras}
\end{equation}
\footnote{One should be careful in implementing commutative limit $\eta \rightarrow 0$; a naive implementation will yield an absurd result, like the limit $\hbar \rightarrow 0$ in the problem of 1D harmonic oscillator having energy spectrum $E_n=(n+\frac{1}{2})\hbar \omega$. The classical continuous spectrum in the limit $\hbar \rightarrow 0$ can be obtained provided we simultaneously take the limit $n \rightarrow \infty$ holding the product $n\hbar$ fixed. Likewise here too, we should take the simultaneous limit $m \rightarrow \infty$ with $\eta \rightarrow 0$ holding their product $m\eta$ to be fixed. This will ensure that we end up with continuous spectrum of a free electron in 3D. Note that here $\eta, m$ play the role of $\hbar$ and $n$ in 1D harmonic oscillator respectively.}
where for spin-up ($s=\frac{1}{2}$), $m=n_1$ and for spin-down ($s=-\frac{1}{2}$), $m=n_1+1$, when 
$n_1$ is the eigenvalue of the number operator $\hat{a}_{1}^{\dagger} \hat{a}_{1}$. 

\section{Degenerate equation of state}
Above description argues for the modification to the available density of states for the 
electrons in the presence of noncommutativity. The Fermi energy $E_F$ of electrons for the $m$th 
level is given by
\begin{equation}
\epsilon_F^2=x_F^2(m) + 1+2m \eta_D,
\end{equation}
where $~\eta_D=\frac{\eta}{m_e^2c^2},~~
\epsilon_F=\frac{E_F}{m_e c^2},~~x_F(m)=\frac{p_F(m)}{m_e c}$.

Due to quantization in the energy levels in the $x-y$ plane, the modified density of state becomes
$\frac{4\pi \eta}{h^3}dp_z$. This can be understood easily from 
equation (\ref{spectras}), by noting that the allowed values of 3-momenta $\vec{p}$ now lie within a cylinder of radius $\sqrt{2m\eta}$, whose axis is along $p_z$ in momentum space.
Hence the electron number density and electron energy density at zero temperature are given by
\begin{equation}\label{densi}
n_e=\sum_{m=0}^{m_{max}} \frac{4\pi m_e^3 c^3 \eta_D}{h^3}g_m x_F(m),
\end{equation}

\begin{equation}\label{densiu}
u=\frac{4\pi m_e^3 c^3 \eta_D}{h^3}\sum_{m=0}^{m_{max}} g_m \int_{0}^{x_F} E(m) dx(m),
\end{equation}
where $g_m$ is the degeneracy such that $g_m=1$ for $m=0$ and $g_m=2$ for $m\ge 1$, 
$m < m_{1}= (\epsilon_F^2-1)/2\eta_D$ and actually in equations 
(\ref{densi}) and (\ref{densiu}), $m$ is taken 
to be the nearest lowest integer of $(\epsilon_F^2-1)/2\eta_D$ for 
every $\epsilon_F$ and $\eta_D$.
Therefore the pressure of the Fermi gas is \cite{pathria,shapiro} :
\begin{eqnarray}
\nonumber
P&=&n_e E_{F}- u\\ \nonumber
 &=&\sum_{m=0}^{m_{max}} \frac{2\pi m_e^4 c^5 \eta_D }{h^3}g_m \Bigg[ \epsilon_{F} x_F(m)\\ 
&-& (1+2m\eta_D) \log \left(\frac{\epsilon_{F}+ x_F(m)}{\sqrt{1+2m\eta_D}}\right)\bigg]. \label{densir} 
\end{eqnarray}

Finally, the mass density is given by
\begin{equation}
\rho=\mu_e m_n n_e,
\label{rho0}
\end{equation}
where $\mu_e$ is the mean molecular weight per electron and $m_n$ is the mass of a neutron.

Now we assume that all the electrons are filled in the lowest Landau level,
hence $m=0$. The validity and condition of this assumption is given at the end of this section, in the subsection ``Discussion".
Now for $m=0$, on using (\ref{densi}) and (\ref{rho0}), we can write the mass density as:
\begin{equation}
\rho =Qx_F(0),
\end{equation}

\begin{equation}
\mbox{where},~~~ Q=\frac{4\pi\mu_e m_nm_e^3c^3}{h^3}\eta_D.
\end{equation}
and EoS reduces to
\begin{equation}
P=\frac{h^3}{8\pi \mu_e^2 m_n^2 m_e^2 c \eta_D } \Bigg[ \rho \sqrt{\rho^2+ Q^2} -Q^2 \log \frac{\rho+ \sqrt{\rho^2+ Q^2}}{Q} \bigg],  \nonumber
\end{equation}

In the case of $x_F(0)>>1$, which corresponds to $\rho^2 >> Q^2$,
EoS further reduces to a simpler polytropic form as
\begin{equation}
P=\frac{h^3}{8\pi \mu_e^2 m_n^2 m_e^2 c \eta_D }\rho^2
\label{aeos}
\end{equation}
where the polytropic index is $n=1$.

However, for the present case, also $x_F^2=\epsilon_F^2-1>2m\eta_D$ which implies
$\epsilon_F^2=2m_1\eta_D+1$ where $0\lesssim m_1<1$, particularly at
the center and for ground level (analog of the lowest Landau level
for the magnetic case) $\sqrt{\epsilon_F^2(0)-1}= x_F(0)=\sqrt{2m_1\eta_D}$ . Therefore,
at center
\begin{equation}
\rho=\rho_c=\frac{4\pi\mu_e m_n m_e^3 c^3}{h^3}\eta_D^{3/2}\sqrt{2m_1},
\label{rho2}
\end{equation}
when $m_1$ can have any value below unity for all electrons to be in the ground level.
On eliminating $\eta_D$ from equations 
(\ref{aeos}) and (\ref{rho2}), one obtains
\begin{eqnarray}
P=K_{nc}\rho^{4/3},\,\,\,{\rm with}\,\,\,K_{nc}=\frac{hc}{2}\left(\frac{m_1}{2\pi\mu_e^4m_n^4}\right)^{1/3}\nonumber \\=1.1\times 10^{15}m_1^{1/3}. 
\label{eos}
\end{eqnarray}
It is to be noted here that the equation of state looks very similar to that of Chandrasekhar except for the fact that the constant of proportionality $K_{nc}$ is not the same. Here it is augmented as compared to the usual case. This will have a bearing on our subsequent results.

\subsection{Fixing noncommutativity parameter}

From equation (\ref{rho2}) with $\rho_c=2\times 10^{10}/V$ g/cc, where 
$V$ is a parameter allowing to change the central density, we can set
$\eta_D$ at the center of the star given by
\begin{eqnarray}
\eta_D=\left(\frac{2\times 10^{10}h^3}{4\pi\mu_e m_n m_e^3 c^3\sqrt{2m_1}V}\right)^{2/3}\approx\frac{456}{(V\mu_e)^{2/3}m_1^{1/3}}.\nonumber \\
\label{eta}
\end{eqnarray} 
Hence, for $\mu_e=2$ (carbon-oxygen white dwarfs) and $V=1$, $\eta_D>287.3$ from equation (\ref{eta}) at center. If $r$ in equation (\ref{cur}) is the 
average separation of any two electrons, at the center with 
$\rho_c\sim 10^{10}$ gm/cc, we obtain $\hbar^2/(m_e c r)^2\sim \eta_D$. This
justifies the noncommutativity under consideration to be spin-dependent curvature induced
noncommutativity, as argued below equation (\ref{cur}). This further clearly
confirms that as density decreases, $r$ increases and, hence, momentum space
tends to become commutative. 
\subsection{Discussion}
The origin of $\eta$ can be traced back from curvature arising in the gravitational interaction between the electrons.
In fact in \cite{Town}, the author has given a similar basic interpretation of momentum NC. The crux of the matter is that whenever momenta do not commute, there has to exist a fundamental length scale, or in other words the commutator determines a scale. This length scale is provided by the inter-electron separation in the given problem which in turn determines the momentum NC parameter $\eta$ through equation (\ref{cur}). 

In case of white dwarf stars, we have $k_B T << E_F$.
This also implies $k_B T << 2 \eta c^2$ because in sufficiently strong values of $\eta$, which is the case here, the energy spacing becomes comparable/greater than electron rest mass energy and the electrons become relativistic i.e., $2\eta c^2 \geq m_e^2 c^4 $.
Here we have, 
 \begin{equation}\label{dimeta}
 \eta_D=\frac{\eta}{ \ m_e^2 c^2} \approx 287 
 \end{equation}
which is quite large as compared to unity.
So, the electrons are ultra-relativistic in our case and this is of utmost interest, for it is this case which is responsible for the existence of mass-limit of white dwarfs even in Chandrasekhar's analysis. Almost all the contribution towards the Fermi energy comes from $\eta$ and so the factor [$(\epsilon_F^2-1)/2\eta_D=m_{max}$] becomes of order unity as evident from (\ref{dimeta}). Now, note that each level specified by $m$ and with a definite Fermi momentum $p_F$ has a large degeneracy (spectrum is dependent on only one mode $n_1$) making the system to be a highly degenerate Fermi system. The case we have here is exactly analogous to that of low temperatures and strong magnetic fields being applied on a material sample; all the electrons are contained within the first few Landau levels. This is exactly in this regime that one encounters the much celebrated quantum Hall effect.

\section{Limiting mass}
Equation (\ref{eos}) is re-written as 
\begin{equation}
P=K_{nc}\rho^{(1+\frac{1}{n})} \label{reos}
\end{equation}
where $n=3$ is the polytropic index.
Following the Lane-Emden formalism, the mass of white dwarfs for 
EoS given by the above equation (\ref{reos}) can be computed \cite{prl} as 
\begin{eqnarray}\label{mass}
M&=&\int_0^R 4\pi r^2\rho dr=4\pi a^3 \rho_c I_n,
\end{eqnarray}
where $\rho=\rho_c\theta^n$, $r=a\xi$ are expressed in terms of dimensionless variables $\theta$ and $\xi$ respectively. Here note that $\rho_c$ is the central density of the white dwarf, and $a$ is given by $a=\sqrt{\frac{(n+1) K_{nc} \rho_c^{(1-n)/n}}{4\pi G}}$.
The radius $R$ of the star is defined as $R=a\xi_1$ when at $\xi=\xi_1$,
$\theta=0$. Note $I_n=\int_0^{\xi_1}\theta^n\xi^2d\xi$.

For $n=3$, $I_n=2.02$ and substituting the value of $K_{nc}$ from (\ref{eos}) in (\ref{mass}) , we get
\begin{eqnarray}
M&=&\left(\frac{h c}{G}\right)^\frac{3}{2} \frac{m_1^{1/2}}{\pi \mu_e^2 m_n^2}I_n=  4.68 M_\odot=M_{max}
\label{mr}
\end{eqnarray}
Hence the mass-limit, when the mass becomes independent of $\rho_c$, turns out to be significantly super-Chandrasekhar for most values of $m_1$ ($0\lesssim m_1<1$), specifically for any $m_1\gtrsim 0.1$ . Here we have taken $\mu_e=2$ which is typically the case for white dwarfs.

\section{Conclusion}
In this paper, we have tried to provide a plausible scenario to explain the recent observations indicating super Chandrasekhar limiting mass. Here we show, albeit heuristically that the gravitational field produced by the stellar material should naturally introduce an effective noncommutativity in momentum space. This noncommutativity can be regarded as a trade of with the curvature effect arising from background gravity in the spirit of \cite{Shahn2}. We then show that this NC in momentum space has some remarkable effect on the EoS and finally on the limiting mass of white dwarf stars, when ultra-relativistic electrons are considered. This can therefore be regarded as a viable proposal to explain this over-luminous peculiar supernova SNeIa. In this context, we would like to mention that it is not
possible to recover the usual Chandrasekhar mass limit
trivially by setting the noncommutativity parameter $\eta$
to be zero. This is because there is a quantization in
the energy levels of an electron stemming from noncom-
mutativity in our analysis, as discussed previously in footnote 1 in page 3, whereas in Chandrasekhar's original anal-
ysis the energy levels are that of a free electron. Further
all of our calculations are based on lowest Landau level
(m=0) which is the extreme quantum case. In order to
get the usual mass limit, one needs to work in the large
m (large quantum number) limit which will then yield
the classical results.

Summing up, the lowest Landau level only filled up case provides the strongest NC contribution in the evaluation of the modified white dwarf mass. Additionally, this case is exactly analytically solved without resorting to numerical computation, which presents the general qualitative features one can have in the present scenario of momentum noncommutativity considered in the paper. To emphasize, note that here the electrons are no more free unlike in Chandrasekhar's work; they are subjected to gravity and this affects the quantum mechanical statistical properties of the electrons. Recently in \cite{psm}, background gravity corrections on the otherwise free electron dynamics inside a white dwarf were studied from a perturbation perspective. Also, it is to be noted that the equation of state (\ref{eos}) looks very similar to that of Chandrasekhar except for the fact that the constant of proportionality is not the same, it has been enhanced. Physically, there is an increase in the degeneracy pressure of the electrons due to the presence of noncommutativity which is responsible for withstanding a greater mass inside the star. Secondly, this is by no means a fully general relativistic treatment of white dwarfs at all. This is because we have ignored the effects of GR when we considered the hydrostatic balance equation for the stellar structure giving rise to the Lane-Emden equations (\ref{mass}). However, Chandrasekhar himself and also others did the full GR computation of the white dwarf mass and there is not much departure from the standard limiting mass without GR \cite{chandras}. In contrast here we find that the momentum scale ($\sqrt{\eta}$), which is ``internally" generated effectively to encapsulate the background gravity on the electrons, is responsible for the considerable modification in the limiting mass of white dwarfs.
This scenario allows for an accessible mass range from Chandrasekhar mass limit all the way upto $4.68 M_\odot$. It is worthwhile to mention here that Chandrasekhar mass limit is an idealization when the radius virtually goes to zero and the density is infinite. All we argue here is that this idealized limit might be enhanced if we have noncommutativity in the momentum components thus providing a plausible theoretical explanation of the recent observations of over-luminous peculiar supernova SNeIa.

Finally recall that the mechanical momentum components in the Landau problem satisfy a  NC algebra where the corresponding NC parameter is proportional to the strength of the external magnetic field \cite{rb,lll}. There, the characteristic frequency is the classical cyclotron frequency given by $\omega_c=eB/m_e c$, where $B$ is the magnetic field strength and 
$e$ is the electron charge. In contrast, here we have the corresponding frequency, $\omega_q=\eta/2m_e\hbar$, as can be computed from (\ref{diagfinal}) and has a purely quantum mechanical origin. Therefore, our analysis may possibly provide a hint at the quantum origin of primordial magnetic field-like effects  \cite{game}. It will be interesting to study the joint effects of momentum space noncommutativity arising from curved space along with a background magnetic field on the fate of Chandrasekhar limit.
\\ \\
\section{acknowledgments}
S.K.P. thanks S. Majid for a useful discussion, constructive criticism and insightful suggestions on the work. S.K.P. thanks Biswajit Chakraborty and Banibrata Mukhopadhyay for suggesting us to look into this avenue and for various advices during the initial stages. P.N. would also like to thank A.A. Deriglazov and Debasish Chatterjee for useful correspondence. The authors would also like to extend their gratefulness to Dept. of Physics, IISC Bangalore for providing hospitality where the work was initiated. We thank the referee for his comments which helped us to improve the manuscript. One of the authors, S.K.P., would like to thank UGC-India for providing financial support in the form of fellowship during
the course of this work.

\end{document}